\documentclass[twoside,twocolumn,9pt]{article}
\usepackage{extsizes}
\usepackage[super,sort&compress,comma]{natbib} 
\usepackage[version=3]{mhchem}
\usepackage[left=1.5cm, right=1.5cm, top=1.785cm, bottom=2.0cm]{geometry}
\usepackage{balance}
\usepackage{mathptmx}
\usepackage{sectsty}
\usepackage{graphicx} 
\usepackage{lastpage}
\usepackage[format=plain,justification=justified,singlelinecheck=false,font={stretch=1.125,small,sf},labelfont=bf,labelsep=space]{caption}
\usepackage{float}
\usepackage{fancyhdr}
\usepackage{fnpos}
\usepackage[english]{babel}
\addto{\captionsenglish}{%
  \renewcommand{\refname}{Notes and references}
}
\usepackage{array}
\usepackage{droidsans}
\usepackage{charter}
\usepackage[T1]{fontenc}
\usepackage[usenames,dvipsnames]{xcolor}
\usepackage{setspace}
\usepackage[compact]{titlesec}
\usepackage{hyperref}
\usepackage{stfloats}

\definecolor{cream}{RGB}{222,217,201}

\begin{document}

\pagestyle{fancy}
\thispagestyle{plain}
\fancypagestyle{plain}{
%%%HEADER%%%
\renewcommand{\headrulewidth}{0pt}
}
%%%END OF HEADER%%%

%%%PAGE SETUP - Please do not change any commands within this section%%%
\makeFNbottom
\makeatletter
\renewcommand\LARGE{\@setfontsize\LARGE{15pt}{17}}
\renewcommand\Large{\@setfontsize\Large{12pt}{14}}
\renewcommand\large{\@setfontsize\large{10pt}{12}}
\renewcommand\footnotesize{\@setfontsize\footnotesize{7pt}{10}}
\renewcommand\scriptsize{\@setfontsize\scriptsize{7pt}{7}}
\makeatother

\renewcommand{\thefootnote}{\fnsymbol{footnote}}
\renewcommand\footnoterule{\vspace*{1pt}% 
\color{cream}\hrule width 3.5in height 0.4pt \color{black} \vspace*{5pt}} 
\setcounter{secnumdepth}{5}

\makeatletter 
\renewcommand\@biblabel[1]{#1}            
\renewcommand\@makefntext[1]% 
{\noindent\makebox[0pt][r]{\@thefnmark\,}#1}
\makeatother 
\renewcommand{\figurename}{\small{Fig.}~}
\sectionfont{\sffamily\Large}
\subsectionfont{\normalsize}
\subsubsectionfont{\bf}
\setstretch{1.125} %In particular, please do not alter this line.
\setlength{\skip\footins}{0.8cm}
\setlength{\footnotesep}{0.25cm}
\setlength{\jot}{10pt}
\titlespacing*{\section}{0pt}{4pt}{4pt}
\titlespacing*{\subsection}{0pt}{15pt}{1pt}
%%%END OF PAGE SETUP%%%

%%%FOOTER%%%
\fancyfoot{}
\fancyfoot[LO,RE]{\vspace{-7.1pt}\includegraphics[height=9pt]{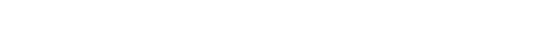}}
\fancyfoot[CO]{\vspace{-7.1pt}\hspace{13.2cm}\includegraphics{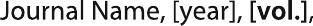}}
\fancyfoot[CE]{\vspace{-7.2pt}\hspace{-14.2cm}\includegraphics{head_foot/RF}}
\fancyfoot[RO]{\footnotesize{\sffamily{1--\pageref{LastPage} ~\textbar  \hspace{2pt}\thepage}}}
\fancyfoot[LE]{\footnotesize{\sffamily{\thepage~\textbar\hspace{3.45cm} 1--\pageref{LastPage}}}}
\fancyhead{}
\renewcommand{\headrulewidth}{0pt} 
\renewcommand{\footrulewidth}{0pt}
\setlength{\arrayrulewidth}{1pt}
\setlength{\columnsep}{6.5mm}
\setlength\bibsep{1pt}
%%%END OF FOOTER%%%

%%%FIGURE SETUP - please do not change any commands within this section%%%
\makeatletter 
\newlength{\figrulesep} 
\setlength{\figrulesep}{0.5\textfloatsep} 

\newcommand{\topfigrule}{\vspace*{-1pt}% 
\noindent{\color{cream}\rule[-\figrulesep]{\columnwidth}{1.5pt}} }

\newcommand{\botfigrule}{\vspace*{-2pt}% 
\noindent{\color{cream}\rule[\figrulesep]{\columnwidth}{1.5pt}} }

\newcommand{\dblfigrule}{\vspace*{-1pt}% 
\noindent{\color{cream}\rule[-\figrulesep]{\textwidth}{1.5pt}} }

\makeatother
%%%END OF FIGURE SETUP%%%

%%%TITLE AND AUTHORS%%%
\twocolumn[
  \begin{@twocolumnfalse}
{\includegraphics[height=30pt]{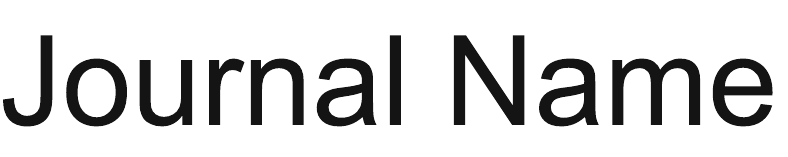}\hfill\raisebox{0pt}[0pt][0pt]{\includegraphics[height=55pt]{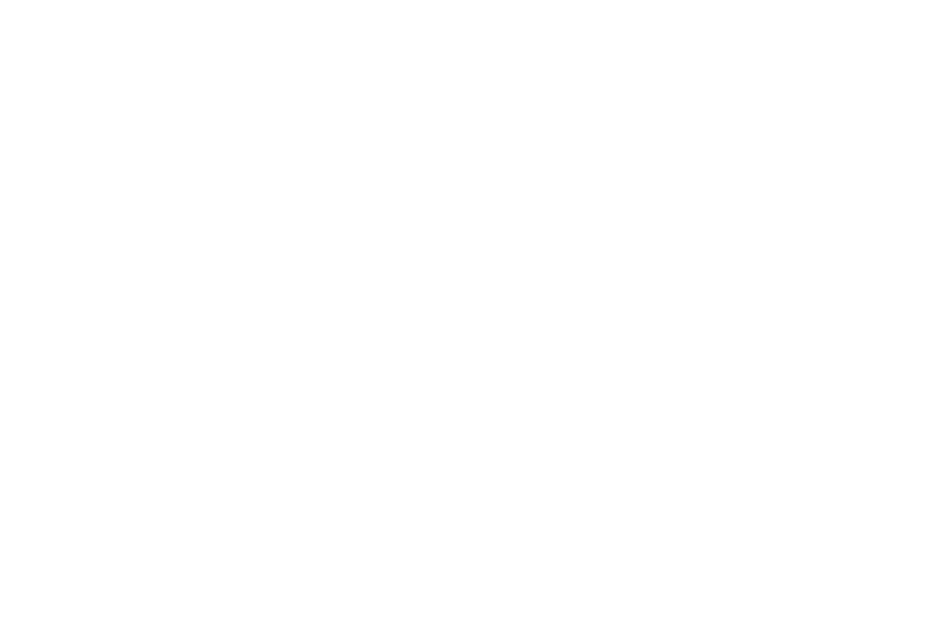}}\\[1ex]
\includegraphics[width=18.5cm]{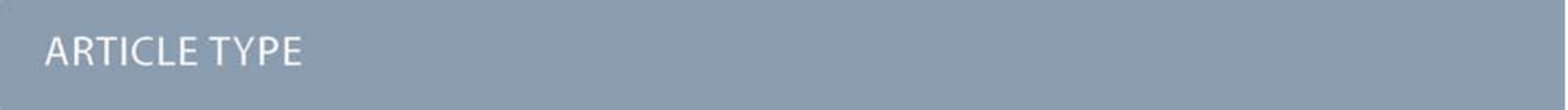}}\par
\vspace{1em}
\sffamily
\begin{tabular}{m{4.5cm} p{13.5cm} }

\includegraphics{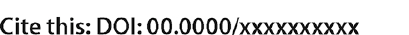} & \noindent\LARGE{\textbf{Remarkably Enhanced Dynamic Oxygen Migration on Graphene Oxide
 Supported by Copper Substrate
}} \\%Article title goes here instead of the text "This is the title"
 & \vspace{0.3cm} \\

 & \noindent\large{Zihan Yan,\textit{$^{a}$}Wenjie Yang,\textit{$^{a}$}Hao Yang,\textit{$^{a}$}Chengao Ji,\textit{$^{a}$} Shuming Zeng,\textit{$^{a}$}Xiuyun Zhang,\textit{$^{a}$}Liang Zhao,$^{\ast}$\textit{$^{a}$} and Yusong Tu$^{\ast}$\textit{$^{a}$}} \\%Author names go here instead of "Full name", etc.

\includegraphics{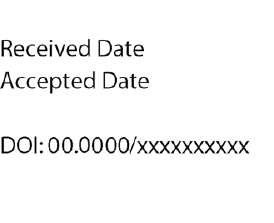} & \\

\end{tabular}

 \end{@twocolumnfalse} \vspace{0.6cm}

  ]
%%%END OF TITLE AND AUTHORS%%%

%%%FONT SETUP - please do not change any commands within this section
\renewcommand*\rmdefault{bch}\normalfont\upshape
\rmfamily
\section*{}
\vspace{-1cm}

%%%FOOTNOTES%%%

\footnotetext{\textit{$^{a}$~College of Physics Science and Technology, Yangzhou University, Jiangsu 225009, China. E-mail: zhaoliang@yzu.edu.cn, ystu@yzu.edu.cn}}

%%%END OF FOOTNOTES%%%

%%%ABSTRACT%%%%

\sffamily{\textbf{The dynamic covalent properties of graphene oxide (GO) are of fundamental interest to a broad range of scientific areas and technological applications. It remains a challenge to access the feasible dynamic reactions for reversibly breaking/reforming covalent bonds of oxygen functional groups on GO, although these reactions can be induced by photonic or mechanical routes, or mediated by adsorbed water. Here, using the density functional theory calculations, we demonstrate the remarkably enhanced dynamic oxygen migration along the basal plane of GO supported by copper substrate (GO@copper), with the \ce{C-O} bond breaking reaction and proton transfer between the neighboring epoxy and hydroxyl groups. Compared to that on GO, the energy barrier of oxygen migration on GO@copper is sharply reduced to be less than or comparable to thermal fluctuations, and meanwhile the crystallographic match between GO and copper substrate induces new oxygen migration paths on GO@copper. This work sheds light on the understanding of metal substrate-enhanced dynamic properties of GO, and evidences the strategy to tune the activity of two-dimension-interfacial oxygen groups for various potential applications.}\\%The abstrast goes here instead of the text "The abstract should be..."

%%%END OF ABSTRACT%%%%

\rmfamily %Please do not remove this line.

%%%MAIN TEXT%%%%

The emergence of dynamic covalent materials for over decades gained tremendous attention in a broad range of scientific areas, such as chemical separation,\cite{hafezi2012adaptation,jin2014dynamic} biological imaging or sensing\cite{zhang2020dynamic} and drug delivery,\cite{zou2017dynamic,rowan2002dynamic,otsuka2013reorganization} especially advances the technological innovations in the manufacturing of materials with novel mechanical and thermal features including self-healing,\cite{zou2017dynamic,zheng2021dynamic,Lehn2015,perera2020dynamic,roy2015dynamers,blaiszik2010self} shape-memory\cite{zheng2021dynamic,perera2020dynamic,zhang2018digital,miao2020demand} and thermoset reprocessing.\cite{zheng2021dynamic,zou2017dynamic} These dynamic behaviors rely on the reversible reactions of breaking/reforming of strong covalent bonds within molecules, endowing materials with adaptivity in response to stimuli and variation of ambient conditions.\cite{imato2012self,chakma2019dynamic,wojtecki2011using,zhu2021unexpected} However, the dynamic reactions usually require harsh conditions and take a long time to reach the thermodynamic equilibrium,\cite{jin2013recent,zhang2020dynamic} which are presumably attributed to the high reaction barrier. Thus, searching for feasible dynamic reactions within a reasonable system is the key to realizing the dynamic covalent materials. Interestingly, we found that mediated by adsorbed water molecules, oxygen groups on graphene oxide (GO) can spontaneously migrate along the basal plane, and GO is converted to a dynamic covalent material with structural adaptivity to the biomolecule adsorption.\cite{tu2020water} However, how to access this dynamic oxygen migration on GO without water remains unclear.

The metal substrate supported GO, as the promising building block of engineering materials and electronic devices, shows superior mechanical strength,\cite{ovid2014metal,xiong2015graphene,hwang2013enhanced,wang2012reinforcement} high electronic and thermal conductivity,\cite{xu2018properties,goyal2012thermal} and sensitive reaction activity.\cite{khalil2018graphene} The presence of metal substrates introduces the metal-GO interaction and the lattice match. In particular, the (111) surface of the copper substrate has been found to assist the epitaxial growth of graphene and shows a better lattice match with the honeycomb plane of GO in experiments.\cite{frank2014interaction,gao2010epitaxial} Here, we performed the density functional theory (DFT) calculations to investigate the dynamic oxygen migrations on GO supported by a copper substrate (GO@copper). 

Unexpectedly, we find that the copper substrate remarkably enhances the dynamic migration of oxygen groups on the basal plane of GO. The energy barrier of oxygen migration is significantly decreased to be less than or comparable to thermal fluctuations. Meanwhile, the crystallographic match between GO and copper substrate induces new oxygen migration paths on GO@copper. To the best of our knowledge, this is the first report of copper-substrate-enhanced dynamic oxygen migrations on GO, which is important for tuning the activity of two-dimension-interfacial oxygen groups for various potential applications.

\section*{Results and discussion}

\begin{figure}[]
  \centering
    \includegraphics[width=0.49\textwidth]{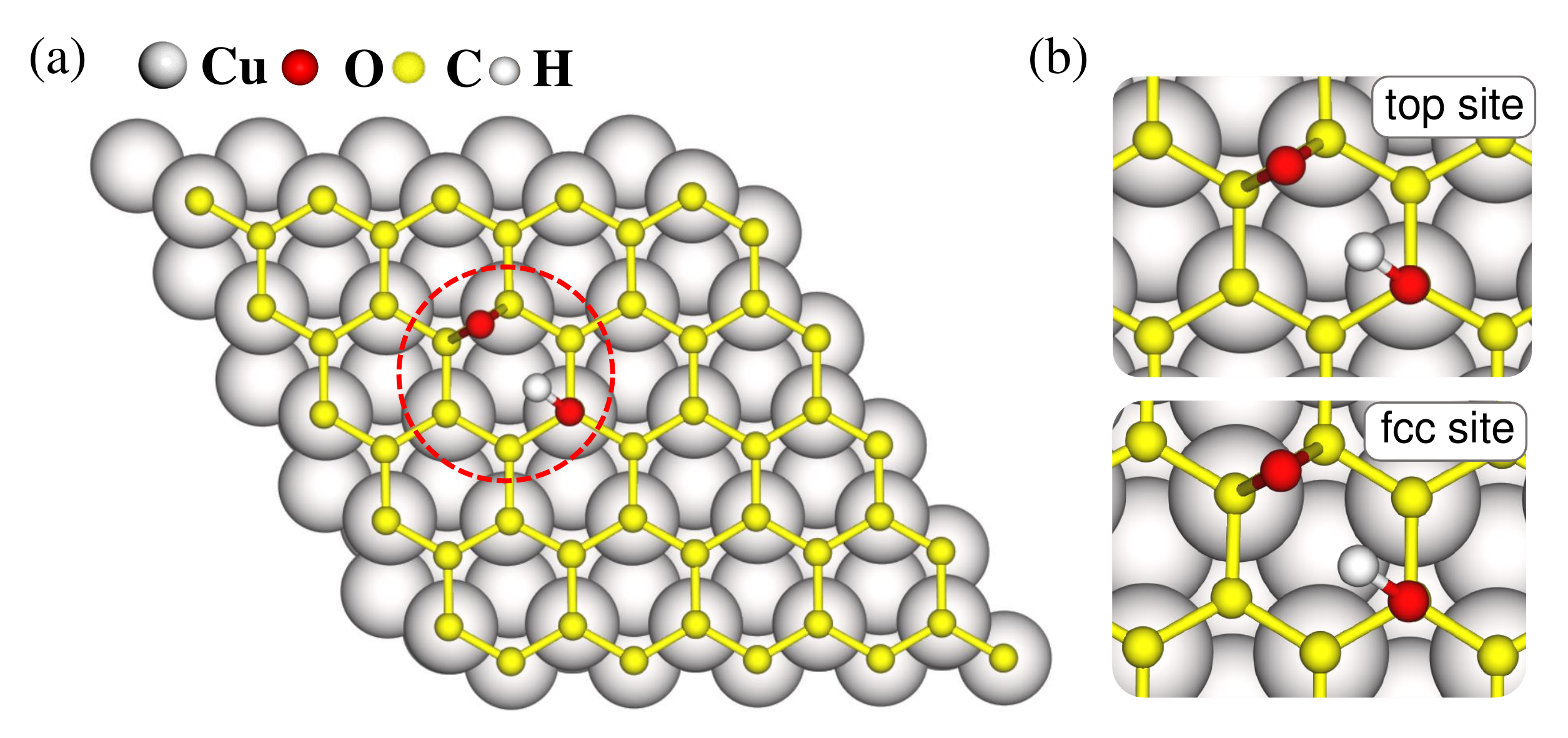}
    \caption{(a) Top view of GO@copper with a pair of hydroxyl and epoxy groups. (b) Zoomed-in snapshots of two different reactant configurations depending on the lattice matching between GO and copper substrate. Top site: Hydroxyl-occupied carbon atom is on the top of a copper atom; Fcc site: Hydroxyl-occupied carbon atom is on the face-centered position of copper atoms.}
    \label{fgr:figure1}
\end{figure}

The dominant oxygen groups on the basal plane of GO, such as hydroxyl and epoxy groups, are distributed with high correlation according to our previous DFT calculations.\cite{yang2014high,tu2020water} As shown in Fig. \ref{fgr:figure1}(a), we choose a pair of hydroxyl and epoxy groups to represent the correlated distribution of oxygen groups. Considering the lattice match between GO and the copper substrate, we denote the hydroxyl-occupied carbon atom on the top of a copper atom and the face-centered position of copper atoms by top site and fcc site, respectively (see Fig. \ref{fgr:figure1}(b)).

\begin{figure*}[bp]
	\centering
	\includegraphics[width=0.7\textwidth]{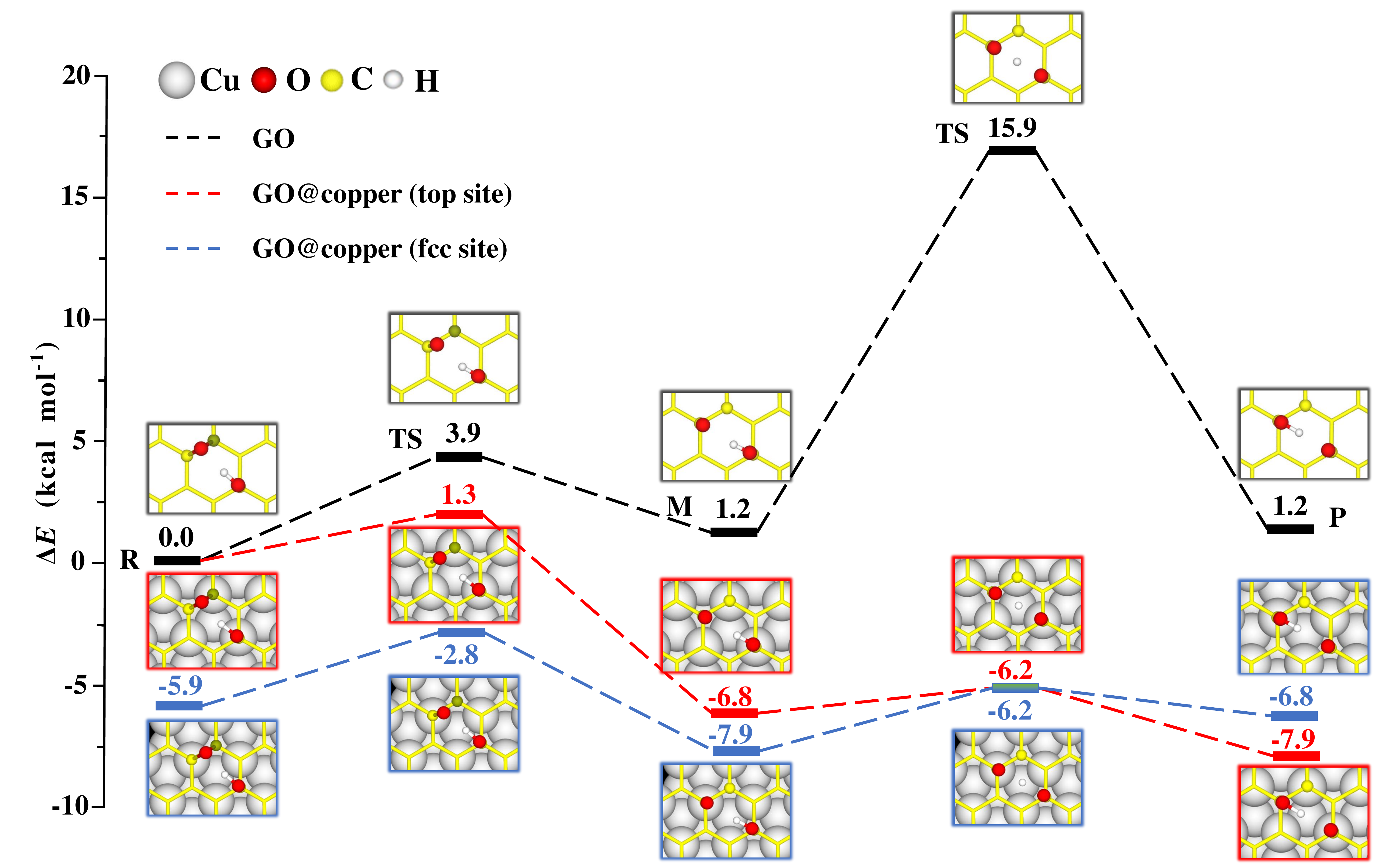}
	\caption{Reaction pathways and state configurations for oxygen migration on GO and GO@copper. The hydroxyl-assisted oxygen migration includes the \ce{C-O} bond breaking reaction and the proton transfer between the dangling oxygen and neighboring hydroxyl for the exchange on GO without the substrate (black lines), on top site of GO@copper (red lines) and fcc site of GO@copper (blue lines). Notations: reactants (R), intermediates (M), transition states (TS), and products (P). The energy levels of R on GO and GO@copper(top site) are shifted for a better comparison.}
	\label{fgr:figure2}
\end{figure*}

The presence of copper substrate significantly decreases the energy barriers of oxygen migration. Figure \ref{fgr:figure2} presents the oxygen migration pathways and state configurations on GO and GO@copper. The oxygen migration assisted by the hydroxyl includes two successive reactions: the \ce{C-O} bond breaking reaction and the proton transfer between the dangling oxygen and neighboring hydroxyl for the exchange. Without the copper substrate, the \ce{C-O} bond breaking reaction has an energy barrier of 3.9 kcal/mol whereas there is a relatively high barrier of 14.7 kcal/mol for the proton transfer, indicating that it is difficult for oxygen groups to migrate spontaneously at ambient conditions.\cite{tu2020water,kumar2014scalable,kim2012room} However, for the top site of GO@copper, the energy barrier for the \ce{C-O} bond breaking reaction reduces to 1.3 kcal/mol. Especially for the proton transfer, we can see a significant decrease of the barrier from 14.7 kcal/mol to 0.6 kcal/mol. These values of energy barriers are lower than or comparable to thermal fluctuations, and also the case for the fcc site (3.1 kcal/mol and 1.7 kcal/mol), suggesting that the migration of oxygen groups can be accessed through the \ce{C-O} bond breaking reaction and proton transfer on GO@copper.

The intermediate (M) is more stable than the reactant (R) due to the presence of the copper substrate. On GO, M is 1.2 kcal/mol higher than R, indicating that the epoxy configuration with two \ce{C-O} bonds is more stable than the dangling \ce{C-O} bond. Whereas on GO@copper, the energy of M is 6.8 kcal/mol (top site) and 2.0 kcal/mol (fcc site) lower than that of R. The lower energy of M with a dangling \ce{C-O} bond indicates that the transition state (TS) with two dangling \ce{C-O} bonds in the proton transfer can also be decreased, which lowers the energy barrier of proton transfer and benefits the spontaneous oxygen migration on GO@copper.

\begin{figure}[]
  \centering
  \includegraphics[width=0.49\textwidth]{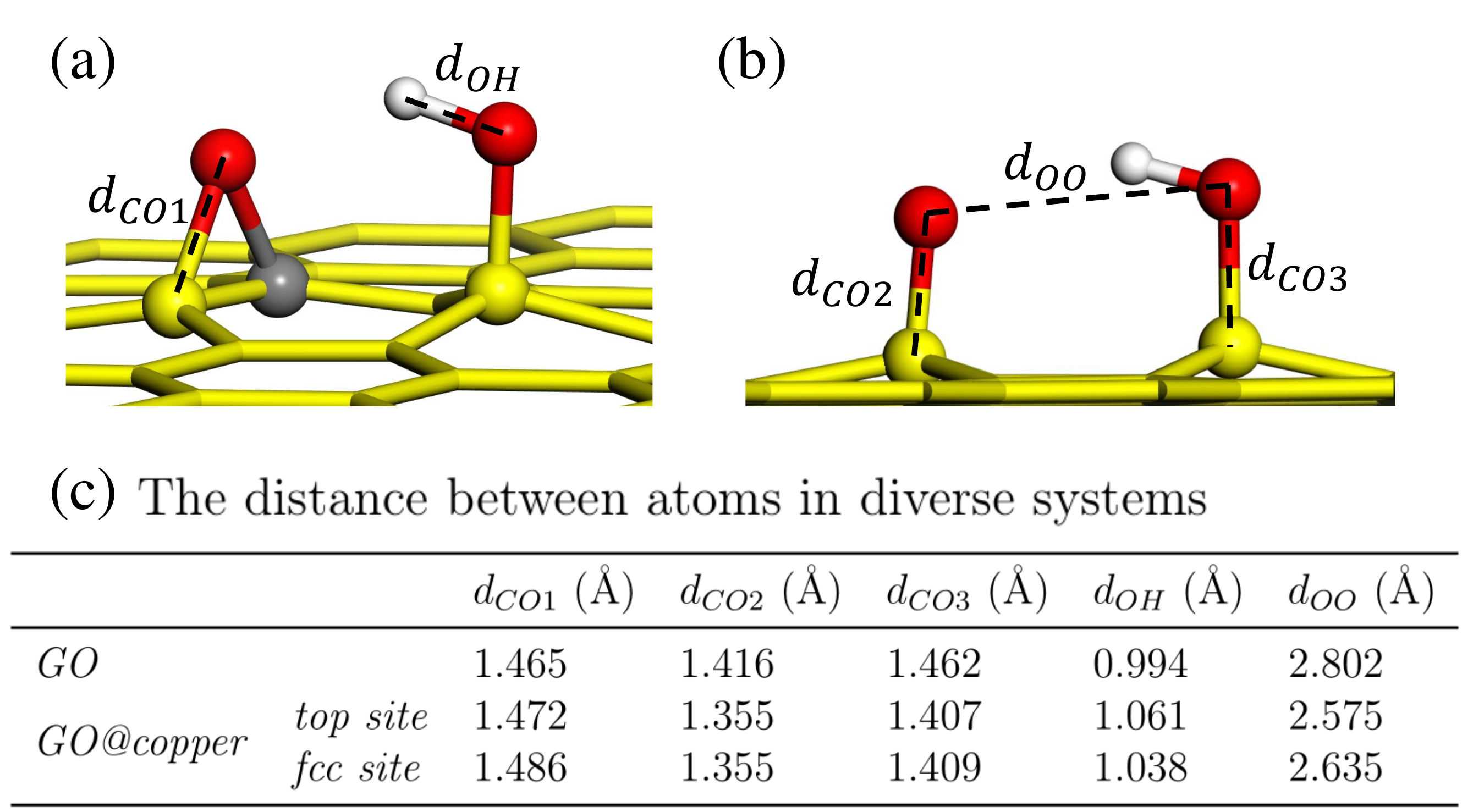}
  \caption{(a, b) Illustrations of distances for \ce{C-O} bonds ($d_{CO1}$, $d_{CO2}$, and $d_{CO3}$), \ce{O-H} bond ($d_{OH}$), and two oxygen atoms ($d_{OO}$) in reactants (R) and intermediates (M). (c) Lists of various distances in R and M on GO, fcc site, and top site of GO@Cu.}
  \label{fgr:figure3}
\end{figure}

Figure \ref{fgr:figure3} compares the variations of bond lengths on GO and GO@copper. Compared to that on GO, the \ce{C-O} bond of epoxy in R on GO@copper is slightly elongated, indicating that the activity of \ce{C-O} bond is enhanced. In M, the \ce{C-O} bonds are significantly shortened, from 1.416 \AA\ to 1.355 \AA\ due to the formation of the dangling \ce{C-O} bond. The shorter \ce{C-O} bond in M on GO@copper suggests the stronger strength of the \ce{C-O} bond, explaining the stability of M with the dangling \ce{C-O} bond. On GO@copper, the \ce{O-O} distance in M becomes shorter and the \ce{O-H} bond is a little elongated, indicating that the proton transfer is more favorable, consistent with the decreased energy barrier of proton transfer on GO@copper.

\begin{figure*}[bp]
  \centering
  \includegraphics[width=0.65\textwidth]{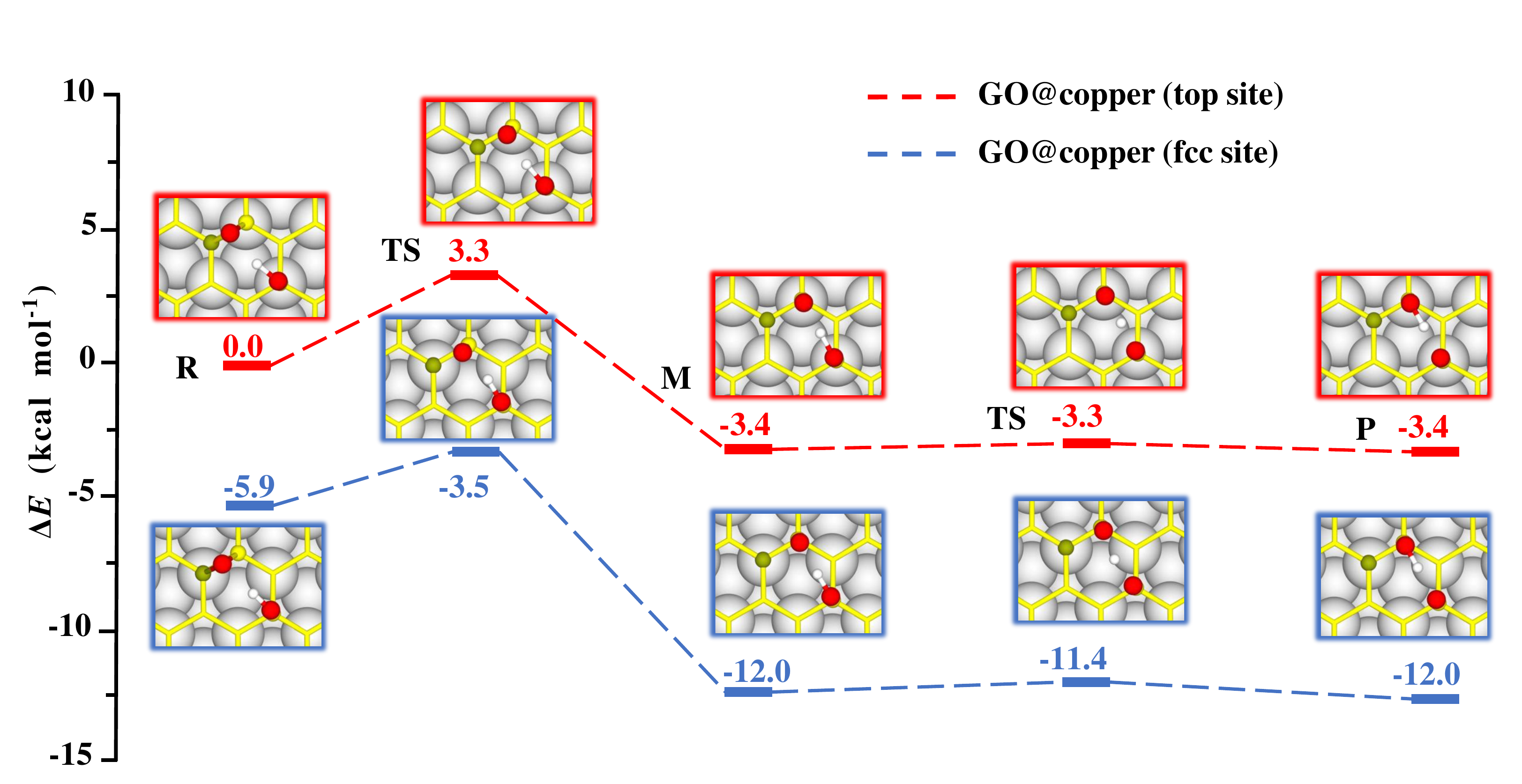}
  \caption{Copper-substrate-induced new reaction pathways of oxygen migration, which occurs at the meta-positions of an aromatic ring, on the top site (blue lines) and fcc site (red lines) of GO@copper.}
  \label{fgr:figure4}
\end{figure*}

The presence of the copper substrate induces new oxygen migration pathways on GO. Generally, the dangling \ce{C-O} bond and the hydroxyl cannot coexist at the meta-position of an aromatic ring of GO due to the steric effects.\cite{yang2014high,tu2020water} However, when GO is supported by the copper substrate, a dangling \ce{C-O} bond is formed at the meta-position of the hydroxyl (see M in Fig. \ref{fgr:figure4}). In this new oxygen migration pathway, a \ce{C-O} bond in epoxy breaks, followed by a proton transfer. The energy barriers for the \ce{C-O} breaking reaction and proton transfer are 3.3 kcal/mol and 0.1 kcal/mol for the top site, 2.4 kcal/mol and 0.6 kcal/mol for the fcc site. These values are all less than or comparable to thermal fluctuations, an indication of feasible migrations of oxygen groups on GO@copper. In particular, the barrier of the proton transfer is only 0.1 kcal/mol, indicating that the proton can transfer rapidly between the dangling oxygen and hydroxyl groups meanwhile the position exchange of these two groups is enabled. We also note that the hydroxyl can migrate within an aromatic ring of GO@copper, via the combination of oxygen migration paths in Figs. \ref{fgr:figure2} and \ref{fgr:figure4}. The new oxygen migration path enriches the understanding of the dynamic behavior of oxygen groups on GO@copper.

\section*{Conclusions}
In summary, we have demonstrated that the copper substrate can remarkably enhance the dynamic oxygen migration on GO@copper based on the DFT calculations. The oxygen migration paths are all accessed by the \ce{C-O} bond breaking reaction and proton transfer between the neighboring epoxy and hydroxyl groups. Compared to that on GO, the energy barrier of oxygen migration on GO@copper is sharply reduced to be less than or comparable to thermal fluctuations. Besides, the crystallographic match between GO and copper substrate induces new oxygen migration paths on GO@copper, where oxygen groups can migrate along the meta-positions of an aromatic ring.

Our work demonstrates the significant dynamic behavior of oxygen groups on GO@copper. The significant dynamic behavior and the enhanced interfacial oxygen activity result from the significantly decreased energy barrier and crystallographic match-induced new oxygen migration paths, which may allow structural adaptation of GO@copper to interfacial oxygen-related reactions. It should be noted that the high-quality and stable growth of GO, and other similar two-dimensional (2D) materials have been fabricated on metal substrates in experiments.\cite{frank2014interaction,gao2010epitaxial,hidalgo2019copper} Therefore, this work evidences the strategy to tune the activity of 2D-interfacial oxygen groups for various potential applications in electrolysis,\cite{Kida2018} capacitors,\cite{Gao2011} fuel cells,\cite{Pandey2017,Gao2014,Karim2013} sensors,\cite{Karim2013} and might provide possibilities for the realization of state-of-the-art high-performance (bio)sensors, biomedical devices, and electronic equipment in the future.

\section*{Method}
The DFT calculations are carried out using the Vienna \textit{Ab-initio} Simulation Package (VASP).\cite{Kresse1994,Kresse1996} The exchange-correlation energy is described by the local density approximation (LDA).\cite{Perdew1981} The convergence accuracy for energy and force in the geometry optimization and transition state (TS) search are 10$^{-5}$ eV and 0.01 eV/\AA, respectively. The projector augmented wave (PAW) is used with the plane wave cutoff of 500 eV. The Monkhorst-Pack\cite{Monkhorst1976} mesh in the k-space is $1 \times 1 \times 1$. The system consists of a GO sheet supported on the (111) surface of the copper substrate of four layers, including 100 copper atoms and 50 carbon atoms, in a periodic $5 \times 5$ supercell. The GO is stretched about 3.7 \% to match the lattice of the copper substrate. The vacuum layer in the \textit{z}-axis direction is set to 15 \AA\ to avoid interactions between periodic images. The copper atoms in the bottom-most layer of the substrate are fixed in all the calculations. The TS is searched via the climbing image-nudged elastic-band (CI-NEB) method.\cite{CINEB2000}

\section*{Author Contributions}
Yusong Tu conceived, designed, and guided the research; Zihan Yan, Wenjie Yang, Hao Yang, and Chengao Ji performed the simulations; Zihan Yan, Shuming Zeng, Liang Zhao, and Yusong Tu analyzed the data; Zihan Yan and Liang Zhao wrote the paper; All the authors participated in discussions of the research.

\section*{Conflicts of interest}
There are no conflicts to declare.

\section*{Acknowledgements}
We are thankful for the helpful suggestions given by PhD candidates Zhijing Huang (Yangzhou University), Guangyu Du (The Hong Kong Polytechnic University), and Xiaoxue Liu (Jilin University). This work was funded by the National Natural Science Foundation of China (Nos. 12075201, 11675138), the Natural Science Foundation of Jiangsu Province (No. BK20201428), the Special Program for Applied Research on Supercomputation of the NSFC-Guangdong Joint Fund (the second phase), and Jiangsu Students’ Innovation and Entrepreneurship Training Program (No. 202111117008Z).
%%%END OF MAIN TEXT%%%

%  For footnotes in the main text of the article please number the footnotes to avoid duplicate symbols. e.g.  \footnote[num]{your text} the corresponding author \ast counts as footnote 1, ESI as footnote 2, e.g. if there is no ESI, please start at [num]=[2], if ESI is cited in the title please start at [num]=[3] etc. Please also cite the ESI within the main body of the text using \dag.

% The \balance command can be used to balance the columns on the final page if desired. It should be placed anywhere within the first column of the last page.

\balance

% If notes are included in your references you can change the title from 'References' to 'Notes and references' using the following command:
\renewcommand\refname{References}

%%%REFERENCES%%%
%\scriptsize{
%\bibliography{ref} %You need to replace "rsc" on this line with the name of your .bib file
%\bibliographystyle{rsc} } %the RSC's .bst file
\providecommand*{\mcitethebibliography}{\thebibliography}
\csname @ifundefined\endcsname{endmcitethebibliography}
{\let\endmcitethebibliography\endthebibliography}{}

\end{document}